\documentstyle[prl,aps,multicol]{revtex}

\renewcommand{\narrowtext}{\begin{multicols}{2} \global\columnwidth20.5pc}
\renewcommand{\widetext}{\end{multicols} \global\columnwidth42.5pc}
\begin{document}
\draft
\title{Universality in invariant random-matrix models: 
Existence near the soft edge}
\author{E. Kanzieper and V. Freilikher}
\address{The Jack and Pearl Resnick Institute of Advanced Technology,\\
Department of Physics, Bar-Ilan University, Ramat-Gan 52900, Israel}
\date{October 21, 1996}
\maketitle

\begin{abstract}
We consider two non-Gaussian ensembles of large Hermitian random matrices
with strong level confinement and show that near the soft edge of the
spectrum both scaled density of states and eigenvalue correlations follow
so-called Airy laws inherent in Gaussian unitary ensemble. This suggests
that the invariant one-matrix models should display universal eigenvalue
correlations in the soft-edge scaling limit.
\end{abstract}

\pacs{\tt chao-dyn/9701005}

\narrowtext

\section{Introduction}

Unitary-invariant random matrix models appear in many physical theories
including nuclear physics, string theory, quantum chaos, and mesoscopic
physics. They are completely defined by the joint distribution function 
\begin{equation}
P\left[ {\bf H}\right] =\frac 1{{\cal Z}_N}\exp \left\{ -%
%TCIMACRO{\limfunc{Tr} }
%BeginExpansion
\mathop{\rm Tr}
%EndExpansion
V\left[ {\bf H}\right] \right\}  \eqnum{1}
\end{equation}
of the entries of the $N\times N$ Hermitian matrix ${\bf H}$. In Eq. (1) the
function $V\left[ {\bf H}\right] $ is referred to as ``confinement
potential'', and it should provide existence of the partition function $%
{\cal Z}_N$. Remarkable feature of this random matrix model is that, under
certain conditions, a particular form of the confinement potential exerts no
influence on the local eigenvalue correlations in the {\it bulk} scaling
limit. More precisely, there is a class of strong even confining potentials $%
V\left( \varepsilon \right) $, increasing at least as fast as $\left|
\varepsilon \right| $ at infinity, for which the two-point kernel in the
bulk of the eigenvalue spectrum follows the {\it sine} form in the large-$N$
limit \cite{Brezin-Zee,Weidenmuller,Kanzieper-1,Kanzieper-2}: 
\begin{equation}
K_{\text{bulk}}\left( s,s^{\prime }\right) =\frac{\sin \left[ \pi \left(
s-s^{\prime }\right) \right] }{\pi \left( s-s^{\prime }\right) }\text{.} 
\eqnum{2}
\end{equation}
This striking property, known as local universality, leads to the conclusion
about universality of arbitrary $n$-point correlation functions $R_n\left(
s_1,...,s_n\right) =\det \left[ K\left( s_i,s_j\right) \right] _{i,j=1...n}$ 
$\left( n>1\right) $ in the local regime. In contrast, the global
characteristics of eigenspectrum, like density of states or one-point
Green's function, display a great sensitivity to the details of confinement
potential \cite{Kanzieper-1}.

Less is known about eigenvalue correlations near the {\it soft edge} which
is of special interest in the matrix models of 2D quantum gravity \cite
{Migdal}. In the early study \cite{Bowick} the behavior of the density of
states near the tail of eigenvalue support has been explored. Authors of
Ref. \cite{Bowick} showed that there is a universal crossover from a
non-zero density of states to a vanishing density of states which is
independent of confining potential in the soft-edge scaling limit. Whereas
the universal behavior of the density of states in the soft-edge scaling
limit has been proven, the (supposed) universality of $n$-point correlations
was not yet considered.

The problem we address in this work is: Whether the eigenvalue correlations
in ensembles of large random matrices also possess a universal behavior in
the soft-edge scaling limit? To provide an answer to this question we first
quote some results for Gaussian unitary ensemble (GUE) that has received the
most study near the soft edge, and then we turn to the consideration of
eigenspectra of two strongly non-Gaussian ensembles of random matrices
associated with quartic and sextic confining potentials.

In the soft-edge scaling limit, GUE is characterized by the {\it Airy}
two-point kernel \cite{Forrester} 
\begin{equation}
K_{\text{GUE}}\left( s,s^{\prime }\right) =\frac{%
%TCIMACRO{\limfunc{Ai} }
%BeginExpansion
\mathop{\rm Ai}
%EndExpansion
\left( s\right) 
%TCIMACRO{\limfunc{Ai} }
%BeginExpansion
\mathop{\rm Ai}
%EndExpansion
^{\prime }\left( s^{\prime }\right) -%
%TCIMACRO{\limfunc{Ai} }
%BeginExpansion
\mathop{\rm Ai}
%EndExpansion
\left( s^{\prime }\right) 
%TCIMACRO{\limfunc{Ai} }
%BeginExpansion
\mathop{\rm Ai}
%EndExpansion
^{\prime }\left( s\right) }{s-s^{\prime }},  \eqnum{3}
\end{equation}
whose spectral properties have got the detailed study in Ref. \cite{Tracy}.
As a consequence of Eq. (3), the scaled density of states, unlike in the
case of bulk scaling limit, cannot already be taken as being approximately
constant, and changes in accordance with Airy law: 
\begin{equation}
\nu _{\text{GUE}}\left( s\right) =\left( \frac d{ds}%
%TCIMACRO{\limfunc{Ai} }
%BeginExpansion
\mathop{\rm Ai}
%EndExpansion
\left( s\right) \right) ^2-s\left[ 
%TCIMACRO{\limfunc{Ai} }
%BeginExpansion
\mathop{\rm Ai}
%EndExpansion
\left( s\right) \right] ^2  \eqnum{4a}
\end{equation}
with asymptotes 
\begin{equation}
\nu _{\text{GUE}}\left( s\right) =\left\{ 
\begin{array}{ll}
\frac{\sqrt{\left| s\right| }}\pi -\frac{\cos \left( 4\left| s\right|
^{3/2}/3\right) }{4\pi \left| s\right| }, & s\rightarrow -\infty , \\ 
\frac 1{8\pi s}\exp \left( -4s^{3/2}/3\right) , & s\rightarrow +\infty .
\end{array}
\right.  \eqnum{4b}
\end{equation}

Our following treatment of non-Gaussian random matrix ensembles with strong
level confinement will be built upon the orthogonal polynomial technique 
\cite{Mehta} allowing to express the two-point kernel for the random-matrix
ensemble defined by Eq. (1) through the polynomials $P_N\left( \varepsilon
\right) $ orthogonal on the whole real axis with respect to the weight $\exp
\left\{ -2V\left( \varepsilon \right) \right\} $. We fix the polynomials $%
P_n $ satisfying the three-term recurrence formula 
\begin{equation}
\varepsilon P_n=a_{n+1}P_{n+1}+a_nP_{n-1}  \eqnum{5}
\end{equation}
to be orthonormal, 
\begin{equation}
\int_{-\infty }^{+\infty }d\varepsilon P_n\left( \varepsilon \right)
P_m\left( \varepsilon \right) \exp \left\{ -2V\left( \varepsilon \right)
\right\} =\delta _{nm}\text{.}  \eqnum{6}
\end{equation}
Under these conditions the two-point kernel reads as 
\begin{equation}
K_N\left( \varepsilon ,\varepsilon ^{\prime }\right) =a_N\frac{\psi _N\left(
\varepsilon ^{\prime }\right) \psi _{N-1}\left( \varepsilon \right) -\psi
_N\left( \varepsilon \right) \psi _{N-1}\left( \varepsilon ^{\prime }\right) 
}{\varepsilon ^{\prime }-\varepsilon }\text{,}  \eqnum{7}
\end{equation}
where \cite{LMS} $a_N=k_{N-1}/k_N$ [$k_N$ is a leading coefficient of the
orthogonal polynomial $P_N\left( \varepsilon \right) $], and the
``wavefunctions'' $\psi _N\left( \varepsilon \right) =P_N\left( \varepsilon
\right) \exp \left\{ -V\left( \varepsilon \right) \right\} $ have been
introduced. Inasmuch as our concern is with the matrices of large
dimensions, $N\gg 1$, the only asymptotics of the ``wavefunctions'' $\psi _N$
are needed, and also a meaningful {\it scaling limit} should be constructed.
Quite generally, this can be done by passing from initial energy variable $%
\varepsilon $ to a new scaled variable $s$ that remains finite as $%
N\rightarrow \infty $: $\varepsilon =\varepsilon \left( N,s\right)
=\varepsilon _s$. Then the scaled two-point kernel is determined by the
formula 
\begin{equation}
K\left( s,s^{\prime }\right) =\lim_{N\rightarrow \infty }K_N\left(
\varepsilon _s,\varepsilon _{s^{\prime }}\right) \frac{d\varepsilon _s}{ds}%
\text{.}  \eqnum{8}
\end{equation}

\section{Quartic confinement potential}

We choose the quartic confinement potential in the form $V\left( \varepsilon
\right) =\frac 12\varepsilon ^4$. In this case the differential equation for 
$\psi _n\left( \varepsilon \right) $ [index $n$ is arbitrary positive
integer] can be obtained by the Shohat's method \cite{Shohat,Nevai}: 
\begin{equation}
\frac{d^2}{d\varepsilon ^2}\psi _n\left( \varepsilon \right) -\left[ \frac d{%
d\varepsilon }\ln \varphi _n\left( \varepsilon \right) \right] \frac d{%
d\varepsilon }\psi _n\left( \varepsilon \right) +Q_n\left( \varepsilon
\right) \psi _n\left( \varepsilon \right) =0,  \eqnum{9a}
\end{equation}
\begin{equation}
\varphi _n\left( \varepsilon \right) =a_{n+1}^2+a_n^2+\varepsilon ^2, 
\eqnum{9b}
\end{equation}
\[
Q_n\left( \varepsilon \right) =\left( 6\varepsilon ^2-4\varepsilon ^6-\frac{%
4\varepsilon ^4}{\varphi _n\left( \varepsilon \right) }\right) +4a_n^2 
\]
\begin{equation}
\times \left( 4\varphi _n\left( \varepsilon \right) \varphi _{n-1}\left(
\varepsilon \right) +1-4a_n^2\varepsilon ^2-4\varepsilon ^4-\frac{%
2\varepsilon ^2}{\varphi _n\left( \varepsilon \right) }\right) .  \eqnum{9c}
\end{equation}
Here $a_n$ is the recursion coefficient entering the corresponding
three-term recurrence formula for the given set of orthogonal polynomials.
Also, the following exact relation takes place: 
\begin{equation}
\psi _{n-1}\left( \varepsilon \right) =\frac{\psi _n^{\prime }\left(
\varepsilon \right) +\psi _n\left( \varepsilon \right) \left[ V^{\prime
}\left( \varepsilon \right) +4\varepsilon a_n^2\right] }{4a_n\varphi
_n\left( \varepsilon \right) }.  \eqnum{10}
\end{equation}

Thereafter we shall be interested in the behavior of the wavefunction $\psi
_n$ near the soft band edge $D_n$ in the limit $n=N\gg 1$. In this case the
endpoint of the spectrum $D_N=2a_N$, where \cite{Lew} 
\begin{equation}
a_N=\left( \frac N{12}\right) ^{1/4}\left[ 1+{\cal O}\left( N^{-2}\right)
\right] ,  \eqnum{11}
\end{equation}
and, 
\begin{equation}
\varphi _N\left( \varepsilon \right) =2a_N^2+\varepsilon ^2+{\cal O}\left(
N^{-1/2}\right) \text{.}  \eqnum{12}
\end{equation}

Let us move the spectrum origin to its endpoint $D_N$, making replacement $%
\varepsilon =D_N+t$, and denote $\widehat{\psi }_N\left( t\right) =\psi
_N\left( \varepsilon -D_N\right) .$ It is straightforward to show that this
function obeys equation 
\begin{equation}
\frac{d^2}{dt^2}\widehat{\psi }_N\left( t\right) -18D_N^5t\cdot \widehat{%
\psi }_N\left( t\right) =0  \eqnum{13}
\end{equation}
in the asymptotic limit $N\gg 1$. When deriving we supposed the
characteristic energy scale $t_v\left( N\right) =$ $\left| d\ln \widehat{%
\psi }_N\left( t\right) /dt\right| ^{-1}$of the variation of $\widehat{\psi }%
_N\left( t\right) $ to be much smaller than the band edge $D_N$.

Solution to Eq. (13) can be written through the Airy function $y\left(
x\right) =%
%TCIMACRO{\limfunc{Ai}}
%BeginExpansion
\mathop{\rm Ai}
%EndExpansion
\left( x\right) $ satisfying the differential equation $y^{\prime \prime
}\left( x\right) -xy\left( x\right) =0$: 
\begin{equation}
\widehat{\psi }_N\left( t\right) =\lambda _N%
%TCIMACRO{\limfunc{Ai} }
%BeginExpansion
\mathop{\rm Ai}
%EndExpansion
\left( t\cdot \left( 18D_N^5\right) ^{1/3}\right) \text{ .}  \eqnum{14}
\end{equation}
One can check that the condition $t_v\left( N\right) \sim {\cal O}\left(
N^{-5/12}\right) \ll D_N$ is fulfilled. The coefficient $\lambda _N$
entering Eq. (14) still remains unknown.

To compute the two-point kernel, Eq. (7), we have to correctly determine the
asymptotic behavior of the $\widehat{\psi }_{N-1}\left( t\right) $. This can
be done by means of the asymptotic analysis of the exact relation Eq. (10),
which in the large-$N$ limit comes down to 
\begin{equation}
\widehat{\psi }_{N-1}\left( t\right) =\widehat{\psi }_N\left( t\right) +%
\frac 1{3D_N^3}\frac d{dt}\widehat{\psi }_N\left( t\right) .  \eqnum{15}
\end{equation}

It is convenient to define the soft-edge scaling limit as 
\begin{equation}
\varepsilon _s=D_N+\frac s{\left( 18D_N^5\right) ^{1/3}}.  \eqnum{16}
\end{equation}
Then the two-point kernel, Eq. (7), and the density of states $K\left(
\varepsilon _s,\varepsilon _s\right) $, are given by the formulas 
\begin{equation}
K\left( \varepsilon _s,\varepsilon _{s^{\prime }}\right) =\lambda _N^2\left( 
\frac 32D_N^4\right) ^{1/3}K_{\text{GUE}}\left( s,s^{\prime }\right) , 
\eqnum{17a}
\end{equation}
and 
\begin{equation}
\nu \left( \varepsilon _s\right) =\lambda _N^2\left( \frac 32D_N^4\right)
^{1/3}\nu _{\text{GUE}}\left( s\right) ,  \eqnum{17b}
\end{equation}
respectively. The latter expression provides a possibility to determine the
unknown constant $\lambda _N$ by fitting the soft-edge density of states,
Eq. (17b), to the bulk density of states \cite{Kanzieper-2} 
\begin{equation}
\nu _{\text{bulk}}\left( \varepsilon _s\right) =\frac{D_N^3}\pi \sqrt{%
1-\left( \frac{\varepsilon _s}{D_N}\right) ^2}\left[ 1+2\left( \frac{%
\varepsilon _s}{D_N}\right) ^2\right]  \eqnum{18}
\end{equation}
taken near the endpoint of the spectrum, Eq. (16), provided $1\ll s\ll
D_N^{5/3}$. Equations (18), (17b), (16), and (4b) yield the value $\lambda
_N^2=\left( 12D_N\right) ^{1/3}.$ Now, making use of the Eqs. (17a), (16),
and (8), we arrive at the following expression for the two-point kernel in
the soft-edge scaling limit: 
\begin{equation}
K_{\text{soft}}\left( s,s^{\prime }\right) =K_{\text{GUE}}\left( s,s^{\prime
}\right) .  \eqnum{19}
\end{equation}

Thus, we conclude that the two-point kernel and the density of states,
computed for the random matrix ensemble with quartic confining potential in
the soft-edge scaling limit, coincide exactly with those for GUE.

\section{Sextic confinement potential}

Now we turn to another ensemble of random matrices which is characterized by
the confinement potential $V\left( \varepsilon \right) =\frac 1{12}%
\varepsilon ^6$. Corresponding wavefunctions $\psi _n$ satisfy the same
differential equation, Eq. (9a), but with \cite{Sheen} 
\[
Q_n\left( \varepsilon \right) =-\frac 14\varepsilon ^{10}+\frac 52%
\varepsilon ^4-\frac 12\varepsilon ^5\left[ \frac d{d\varepsilon }\ln
\varphi _n\left( \varepsilon \right) \right] 
\]
\[
+a_n^2\varphi _n\left( \varepsilon \right) \varphi _{n-1}\left( \varepsilon
\right) -\left( \varepsilon ^5+\pi _n\left( \varepsilon \right) -\frac d{%
d\varepsilon }\right) \pi _n\left( \varepsilon \right) 
\]
\begin{equation}
-2\varepsilon \frac{\pi _n\left( \varepsilon \right) }{\varphi _n\left(
\varepsilon \right) }\left( 2\varepsilon ^2+a_n^2+a_{n+1}^2\right) , 
\eqnum{20a}
\end{equation}
\begin{equation}
\pi _n\left( \varepsilon \right) =a_n^2\varepsilon \left(
a_{n-1}^2+a_n^2+a_{n+1}^2+\varepsilon ^2\right) ,  \eqnum{20b}
\end{equation}
and 
\[
\varphi _n\left( \varepsilon \right) =a_{n+1}^2\left(
a_{n+2}^2+a_{n+1}^2+a_n^2\right) 
\]
\begin{equation}
+a_n^2\left( a_{n+1}^2+a_n^2+a_{n-1}^2\right) +\varepsilon ^2\left(
a_{n+1}^2+a_n^2+\varepsilon ^2\right) \text{.}  \eqnum{20c}
\end{equation}
Also, the following relationship holds for arbitrary $n$: 
\begin{equation}
\psi _{n-1}\left( \varepsilon \right) =\frac{\psi _n^{\prime }\left(
\varepsilon \right) +\psi _n\left( \varepsilon \right) \left[ V^{\prime
}\left( \varepsilon \right) +\pi _n\left( \varepsilon \right) \right] }{%
a_n\varphi _n\left( \varepsilon \right) }.  \eqnum{21}
\end{equation}

The asymptotic analysis of the solution to the second-order differential
equation near the endpoint of the spectrum is quite similar to that done in
preceding section. Therefore, we only sketch its main points.

For $n=N\gg 1$ the recursion coefficient \cite{MN} 
\begin{equation}
a_N=\left( \frac N{10}\right) ^{1/6}\left[ 1+{\cal O}\left( N^{-2}\right)
\right] ,  \eqnum{22}
\end{equation}
and 
\begin{equation}
\varphi _N\left( \varepsilon \right) =6a_N^4+\varepsilon ^2\left(
\varepsilon ^2+2a_N^2\right) +{\cal O}\left( N^{-1/2}\right) ,  \eqnum{23}
\end{equation}
\begin{equation}
\pi _N\left( \varepsilon \right) =a_N^2\varepsilon \left( \varepsilon
^2+3a_N^2\right) +{\cal O}\left( N^{-1/3}\right) .  \eqnum{24}
\end{equation}
Introducing the shifted energy variable, $\varepsilon =D_N+t$, we are able
to rewrite the differential equation (9a) for the function $\widehat{\psi }%
_N\left( t\right) =\psi _N\left( \varepsilon -D_N\right) $ in the form 
\begin{equation}
\frac{d^2}{dt^2}\widehat{\psi }_N\left( t\right) -\frac{225}{128}D_N^9t\cdot 
\widehat{\psi }_N\left( t\right) =0,  \eqnum{25}
\end{equation}
assuming that the characteristic energy scale $t_v\left( N\right) =$ $\left|
d\ln \widehat{\psi }_N\left( t\right) /dt\right| ^{-1}$of the variation of $%
\widehat{\psi }_N\left( t\right) =\psi _N\left( \varepsilon -D_N\right) $ is
much smaller than the band edge $D_N$.

Solution of Eq. (25) takes the form 
\begin{equation}
\widehat{\psi }_N\left( t\right) =\lambda _N%
%TCIMACRO{\limfunc{Ai} }
%BeginExpansion
\mathop{\rm Ai}
%EndExpansion
\left( \left( \frac{225D_n^9}{128}\right) ^{1/3}\,t\right) ,  \eqnum{26}
\end{equation}
with coefficient $\lambda _N$ that will be determined later by the same
fitting arguments. The assumption $t_v\left( N\right) \ll D_N$ is obviously
fulfilled.

To get the asymptotic behavior of $\widehat{\psi }_{N-1}\left( t\right) $ in
the large-$N$ limit, we simplify Eq. (21) to 
\begin{equation}
\widehat{\psi }_{N-1}\left( t\right) =\widehat{\psi }_N\left( t\right) +%
\frac{16}{15D_N^5}\frac d{dt}\widehat{\psi }_N\left( t\right) .  \eqnum{27}
\end{equation}

It is convenient to define the soft-edge scaling limit as 
\begin{equation}
\varepsilon _s=D_N+\frac s{D_N^3}\left( \frac{128}{225}\right) ^{1/3}\text{.}
\eqnum{28}
\end{equation}
Then the two-point kernel is 
\begin{equation}
K\left( \varepsilon _s,\varepsilon _{s^{\prime }}\right) =\lambda
_N^2D_N^2\left( \frac{15}{32}\right) ^{1/3}K_{\text{GUE}}\left( s,s^{\prime
}\right)  \eqnum{29a}
\end{equation}
while the density of states takes the form 
\begin{equation}
\nu \left( \varepsilon _s\right) =\lambda _N^2D_N^2\left( \frac{15}{32}%
\right) ^{1/3}\nu _{\text{GUE}}\left( s\right) .  \eqnum{29b}
\end{equation}

The fitting arguments, based on the expansion of the Eq. (29b) and of the
bulk density of states \cite{Kanzieper-2} 
\[
\nu _{\text{bulk}}\left( \varepsilon _s\right) =\frac{D_N^5}{16\pi }\sqrt{%
1-\left( \frac{\varepsilon _s}{D_N}\right) ^2} 
\]
\begin{equation}
\times \left[ 3+4\left( \frac{\varepsilon _s}{D_N}\right) ^2+8\left( \frac{%
\varepsilon _s}{D_N}\right) ^4\right]  \eqnum{30}
\end{equation}
near the soft edge [when $1\ll s\ll D_N^3$], yield $\lambda _N^2=D_N\left(
15/4\right) ^{1/3}.$ Combining Eqs. (29a), (28), and (8), we end with the
following expression for the two-point kernel in the soft-edge scaling
limit: 
\begin{equation}
K_{\text{soft}}\left( s,s^{\prime }\right) =K_{\text{GUE}}\left( s,s^{\prime
}\right) .  \eqnum{31}
\end{equation}

This formula demonstrates that in the soft-edge scaling limit the eigenlevel
properties for the random matrix ensemble with sextic confinement potential
are determined by the same Airy law which is inherent in GUE.

\section{Concluding remarks}

We have considered for the first time the correlations of the eigenlevels
near the soft edge for two strongly non-Gaussian ensembles of large random
matrices possessing unitary symmetry, and associated with quartic and sextic
confinement potentials. Our treatment has been based on the analysis of the
second-order differential equations for the corresponding ``wavefunctions''
near the soft edge. In both cases it was found that correlations between
appropriately scaled eigenvalues are universal, and characterized by the
Airy two-point kernel Eq. (3) which previously has been found for GUE.

Together with the fact of universal behavior of the density of states,
previously proven in Ref. \cite{Bowick}, the consideration presented gives a
strong impression that spectral correlations in invariant ensembles of large
random matrices with rather strong and monotonous confinement potential are
indeed universal near the soft edge.

\begin{center}
{\bf Acknowledgment}
\end{center}

The authors thank Prof. Craig A. Tracy for bringing this problem to our
attention. The support of the Ministry of Science of Israel (E. K.) is
gratefully acknowledged.

\widetext

\end{document}